\documentclass[aps,twocolumn,prl]{revtex4-1}
\usepackage{amsmath,amssymb,graphicx,epsfig}
\usepackage[english]{babel}
\usepackage{epstopdf}

\newcommand{\Iota}{\mathrm{I}}
\newcommand{\Tau}{\mathrm{T}}
\newcommand{\nobracket}{}
\newcommand{\nocomma}{}
\newcommand{\nosymbol}{}

\newcommand{\tmop}[1]{\ensuremath{\operatorname{#1}}}

\newcommand{\tmrsup}[1]{\textsuperscript{#1}}
\newcommand{\tmtextit}[1]{{\itshape{#1}}}

\begin{document}

\title{Dynamical Coulomb Blockade of Shot Noise}

\author{Carles Altimiras}
\altaffiliation{Presently at NEST, Istituto Nanoscienze-CNR and Scuola Normale
Superiore, I-56127 Pisa, Italy}

\author{Olivier Parlavecchio}

\author{Philippe Joyez}

\author{Denis Vion}

\author{Patrice Roche}

\author{Daniel Esteve}

\author{Fabien Portier}
\email{fabien.portier@cea.fr}
\affiliation{Service de Physique de l'Etat Condens{\'e} (CNRS URA 2464),
IRAMIS, CEA-Saclay, 91191 Gif-sur-Yvette, France}

\date{September 5, 2013\\
}

\begin{abstract}
  We observe the suppression of the finite frequency shot-noise produced by a
  voltage biased tunnel junction due to its interaction with a 
  single electromagnetic mode of high impedance. The tunnel junction is embedded in a $\lambda
  /4$ resonator containing a dense SQUID array providing it with a
  characteristic impedance in the k$\Omega$ range and a resonant frequency
  tunable in the 4-6 GHz range. Such high impedance gives rise to a sizeable
  Coulomb blockade on the tunnel junction ($\sim$30\% reduction in the
  differential conductance) and allows an efficient measurement of the
  spectral density of the current fluctuations at the resonator frequency. 
  The observed blockade of shot-noise is found in agreement with an
  extension of the dynamical Coulomb blockade theory.
\end{abstract}

{\maketitle}

Contrarily to usual electronic components for which one can define an intrinsic behavior (e.g. the I-V characteristic), the transport properties of a coherent quantum conductor depend on its biasing circuit. This is true even when the size of the circuit exceeds the electron coherence length, suppressing electronic interference effects. This non-intrinsic behavior can be traced to the quantum-probabilistic character of the transmission of electrons through the conductor, resulting in broad-band fluctuations of the current called shot noise {\cite{Blanter20001}}. This current noise can create
collective excitations (hereafter called ``photons'') in the electromagnetic
environment seen by the conductor. This yields a back action on the transport properties of the conductor itself
{\cite{ingoldnazarov1992DCB}}. This physics bears similarities with the
spontaneous emission of photons by an excited atom, albeit with important
differences: first, dc biased quantum conductors are out of equilibrium open
systems and cannot be described as a set of discrete levels. Second, the
dimensionless parameter characterizing the electron-photon coupling is given
by the ratio of the environment's impedance to the resistance quantum
$R_{\text{K}} =h/e^{2} \simeq 25.8  \text{k} \Omega$; Hence, by increasing
the impedance of the circuit connected to the quantum conductor, one can
significantly increase the effective coupling constant. This results in a rich
physics, already partially understood: noticeably, the dynamical Coulomb
Blockade (DCB) theory {\cite{ingoldnazarov1992DCB}} accounts for the observed
suppression
{\cite{ClelandDCBinTunnelPRB1992,PhysRevLett.63.1180,0295-5075-10-1-014}} of
the low voltage conductance of a tunnel element as a result of its coupling to
a dissipative electromagnetic environment. A natural step is then to
understand how the coupling to the environment modifies the current
fluctuations themselves: is there a Coulomb Blockade of shot
noise? This question of current fluctuations in the presence of DCB was
addressed theoretically for the low frequency/long time limit, where the
corrections to the noise power and to the full counting statistics were
predicted
{\cite{GalaktionovPRB68p085317,KindermannPRL91p136802,KindermannPRB69p035336,SafiPRL93p126602}}.
Instead, we consider here the environment feedback on the frequency dependence
of the shot noise of a simple quantum conductor, a tunnel junction. Following
Ref. {\cite{PhysRevB.53.7383}}, we extend the DCB theory to predict
the finite frequency emission noise spectrum of a voltage biased tunnel
junction in the presence of an arbitrary linear environment. Probing this prediction
requires to achieve strong coupling of the junction to its environment and to
measure its high frequency shot noise. To do so, we fabricate a tunnel
junction embedded in the simplest environment, a harmonic oscillator, and
measure the effect of Coulomb blockade on the shot noise power at the
frequency of the oscillator. The oscillator is realized with a microwave
resonator based on a Josephson transmission line allowing both a tenfold
increase of the coupling constant between the junction and the resonator, and
to tune the resonant frequency. The data are found in quantitative agreement
with the theory.

In order to evaluate the current and its fluctuations through a tunnel
element in the presence of DCB, we consider a circuit consisting of a tunnel
junction of conductance $G_{\Tau}$ in series with an impedance $Z ( \nu )$
described as the series combination of harmonic modes (see upper panel of Fig.
1) at  temperature $T$ and biased at voltage $V$. We then
compute (see the Supplemental Material for more details) the current $I$
and the quantum spectral density $S_{\Iota} ( \nu )$ of current noise, i.e.
the Fourier transform of the non-symmetrized current-current correlator
\begin{equation}
  S_{\Iota} ( \nu ) =2 \int_{- \infty}^{\infty} \langle I(t)I(0) \rangle
  e^{-i2 \pi \nu t} dt. \label{eq:SII}
\end{equation}
In this convention positive (resp. negative) frequencies correspond to energy
being emitted (resp. absorbed) by the quasiparticles to (resp. from) the
electromagnetic modes. Taking separate thermal equilibrium averages over the
unperturbed quasiparticle and environmental degrees of freedom yields
\begin{eqnarray*}
  I (V) & = & \frac{G_{\text{T}}}{e}  [ \nobracket \gamma \ast P(eV)- \gamma
  \ast P( -eV)], \\
  S_{\Iota} ( \nu ,V) & = & 2G_{\text{T}}  [ \gamma \ast P(eV-h \nu )+ \gamma
  \ast P( -h \nu -eV)], 
\end{eqnarray*}
where $\gamma \ast P (E) = \int d \varepsilon'
\gamma ( \varepsilon' ) P (E- \epsilon' )$ with $P ( \varepsilon )$ the
probability density for a tunneling electron to emit the energy $\varepsilon$
in form of photons into the impedance {\cite{ingoldnazarov1992DCB}}, with
$\gamma ( \epsilon ) = \int d \varepsilon' f ( \varepsilon' )  [1-f(
\varepsilon' + \varepsilon )] = \varepsilon / (1-e^{- \varepsilon /k_{B} T}
)$, and with $f$ the Fermi function. Eq. 2 is the standard DCB expression for 
the tunneling current {\cite{ingoldnazarov1992DCB}}, whereas Eq. 3
is our prediction for the Coulomb Blockade of shot noise, which we probe in
the experiment described below. For a positive bias voltage and low
temperature $(k_{\text{B}} T \ll eV,h \nu )$, Eqs. 2-3 take the
simpler form
\begin{eqnarray}
  I (V) & = & \frac{G_{\Tau}}{e}  \int_{0^{\nosymbol}}^{eV} (eV- \varepsilon )
  P ( \varepsilon ) d \varepsilon , \\
  S_{\Iota} ( \nu ,V) & = & 2G_{\Tau} \Theta ( \tmop{eV} -h \nu ) \!\!
  \int_0^{eV-h \nu} \mkern-18mu (eV-h\nu - \varepsilon) P (\varepsilon)
  d \varepsilon , 
\end{eqnarray}
with $\Theta ( \varepsilon )$ the Heavyside function. Eqs. 4-5 are easily
interpreted: the total current is proportional to the average energy available
for quasiparticles upon the transfer of an electron through the circuit, and
so is the noise power at frequency $\nu$, albeit imposing the emission of a
photon of energy $h \nu$ into the environment. In the case of a vanishing
impedance $Z ( \nu ) \ll R_{\text{K}}$, $P ( \varepsilon ) = \delta (
\varepsilon )$ and one recovers the standard, non interacting, finite
frequency shot noise result \cite{LesovikLoosen}. In the case of a discrete harmonic oscillator of
frequency $\nu_{0} =1/ \left[ 2  \pi   \sqrt{L C} \right]$ and impedance
$Z_{C} = \sqrt{L/C}$, thermalized at a temperature $T  \ll  h  \nu_{0}
/k_{\text{B}}$: $P ( E ) = \sum^{\infty}_{k=0} p_{k}   \delta ( E-k h \nu_{0}
) \nocomma   \nocomma$, with $p_{k} =e^{- \alpha}   \alpha^{k} /k!$ the
probability for the oscillator to absorb $k$ photons
{\cite{ingoldnazarov1992DCB}}, and $\alpha = \pi Z_{C} /R_{K}$ the coupling
strength between the tunnel junction and the oscillator. Our experiment
achieves an unprecedented electron-single mode coupling $\alpha \sim 0.3$, which
allows observing multi-photon processes both in the average current and in the
emission noise. Note that, despite similar denominations, the effect we consider here differs from \textit{static} 
Coulomb Blockade, which results from the charging energy of a small island connected to reservoirs by tunnel barrier.
Static Coulomb Blockade is a quasiclassical effect which can be described by master rate equations, at the level of the current noise \cite{KafanovPhysRevB80}, and even the full counting statitsics \cite{GustavssonPhysRevB74}.

\begin{figure}[h]
  \resizebox{8cm}{!}{\includegraphics{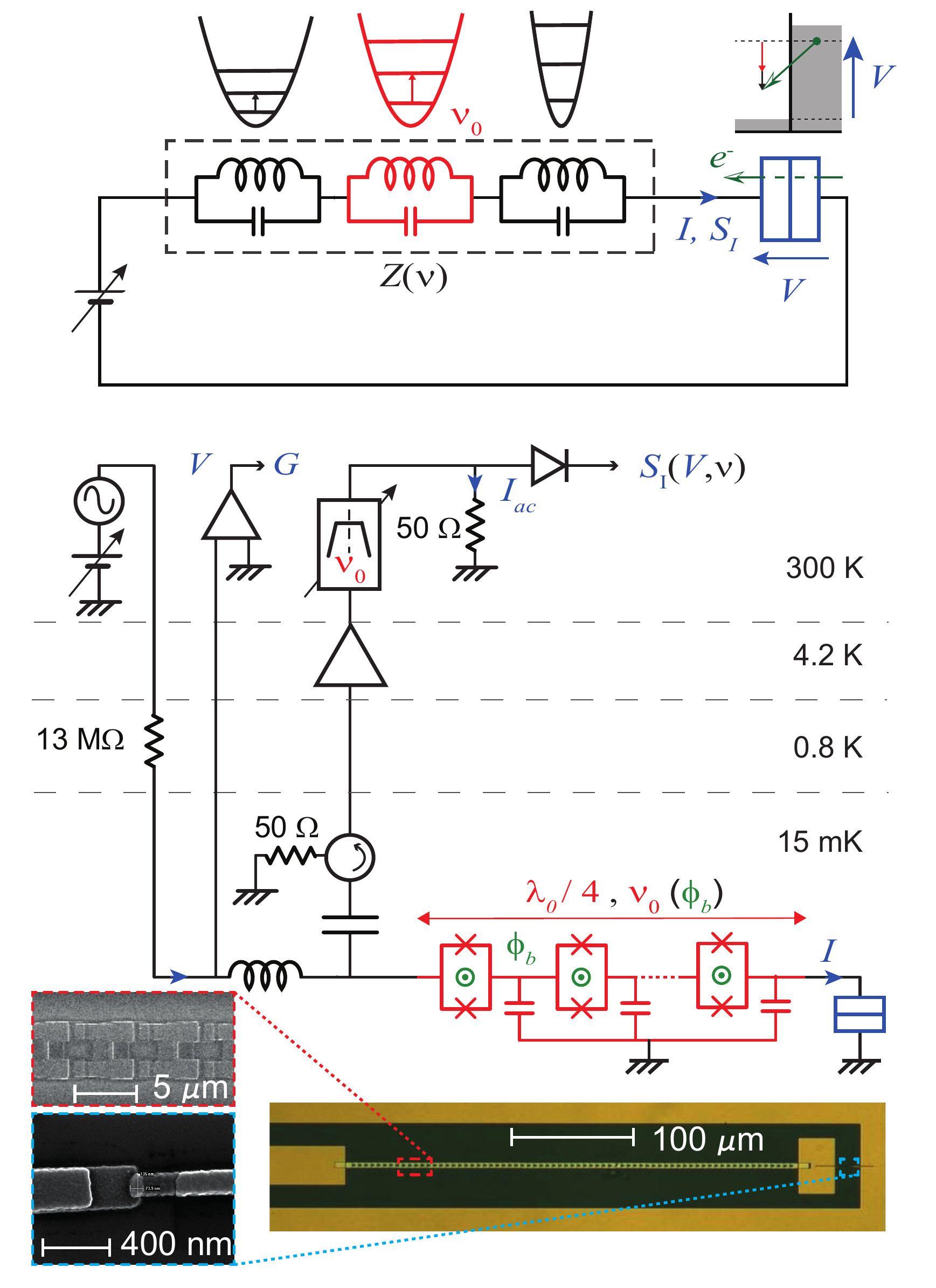}}
  \caption{ Coulomb Blockade in a normal quantum conductor: (a)  A quantum conductor (here a tunnel junction)
  is voltage biased $(V)$ through a series impedance $Z$ modeled as a
  collection of harmonic modes, resulting in inelastic electron tunneling. (b) Experimental set-up: a tunnel junction
  is dc biased at voltage $V$, and  connected to
  a SQUID-based resonator presenting a discrete mode at frequency $\nu_{0}$,  tuned by varying the
  magnetic flux $\phi_{b}$ threading each SQUID loop.
  The dc biasing line and
  $50 \Omega$ microwave measurement line are separated by a bias tee, allowing to measure the
  junction dc differential conductance $G(V)$ and the emission noise $S_I(\nu_0,V)$. The
  measurement line includes an isolator, a cryogenic amplifier with 42 dB
  gain, a 180 MHz passband filter centered on $\nu_{0}$, and a matched
  quadratic detector. Temperatures of the
  different stages are indicated on the right. Bottom illustrations: global view of the sample, with SEM pictures of 
	SQUIDs (top inset) forming the array and of the normal tunnel junction (bottom inset), both from sample 1}
\end{figure}

\begin{figure}[h]
\centering
\includegraphics[width= 9 cm]{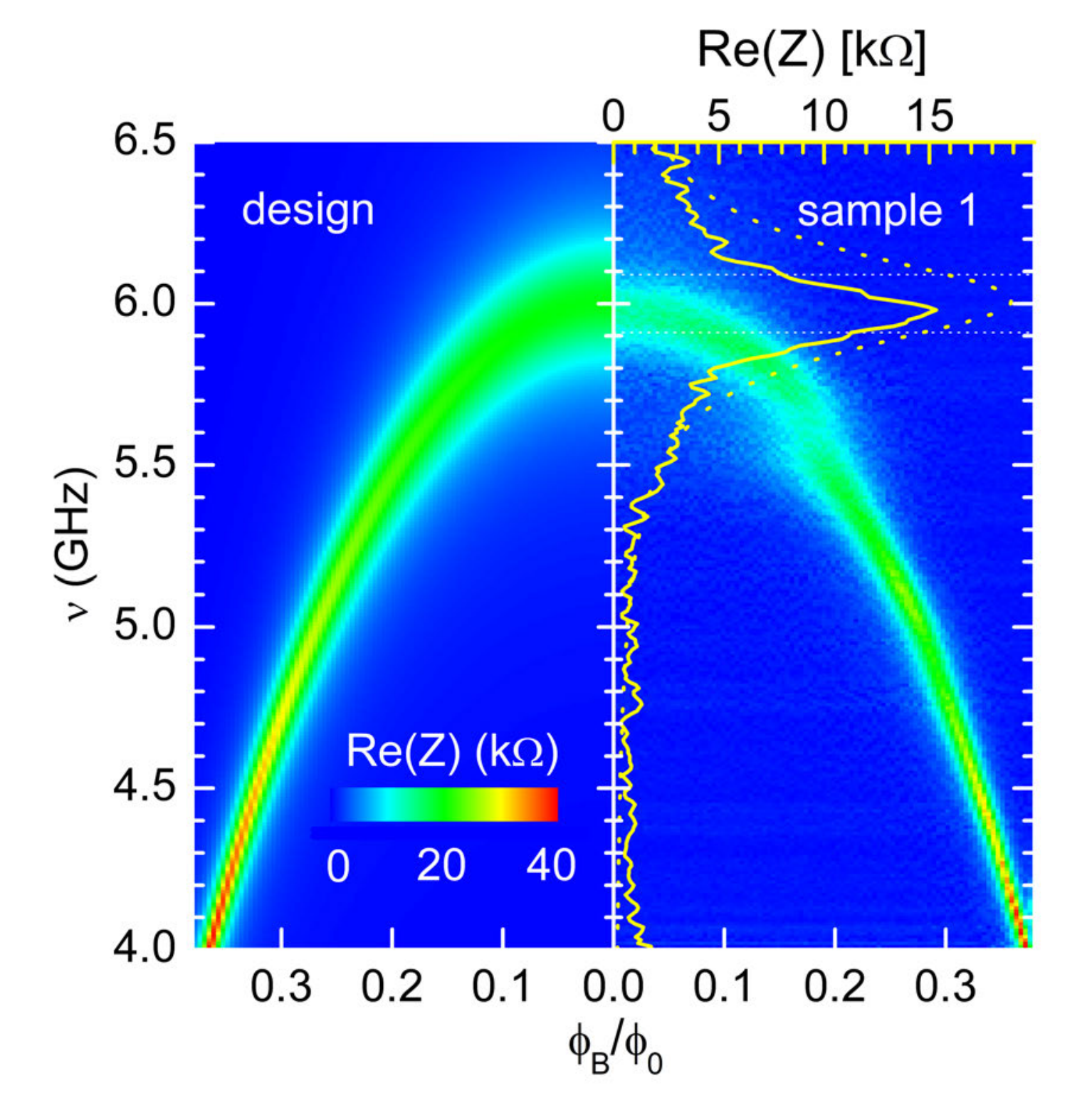}
    \caption{Characterization of the environment impedance:
    Designed (left) and measured (right) real part of the impedance $Z ( \nu
    )$ of the quarter-wave resonator of sample 1 (see Fig.1) as a function of
    magnetic flux $\phi_{b}$ and frequency $\nu$. The overprinted curve on the
    right (top scale) shows the resonance at $\nu_{0} \backsimeq 6
    \mathrm{\tmop{GHz}}$ for $\phi_{b} =0$, measured (solid line) and calculated (dotted line). Horizontal dotted white
    lines indicate -3dB bandwidth used for
    measuring the shot noise power shown in Fig 3a.}
\end{figure}

\begin{figure*}
\centering
\includegraphics[width=1.0\textwidth]{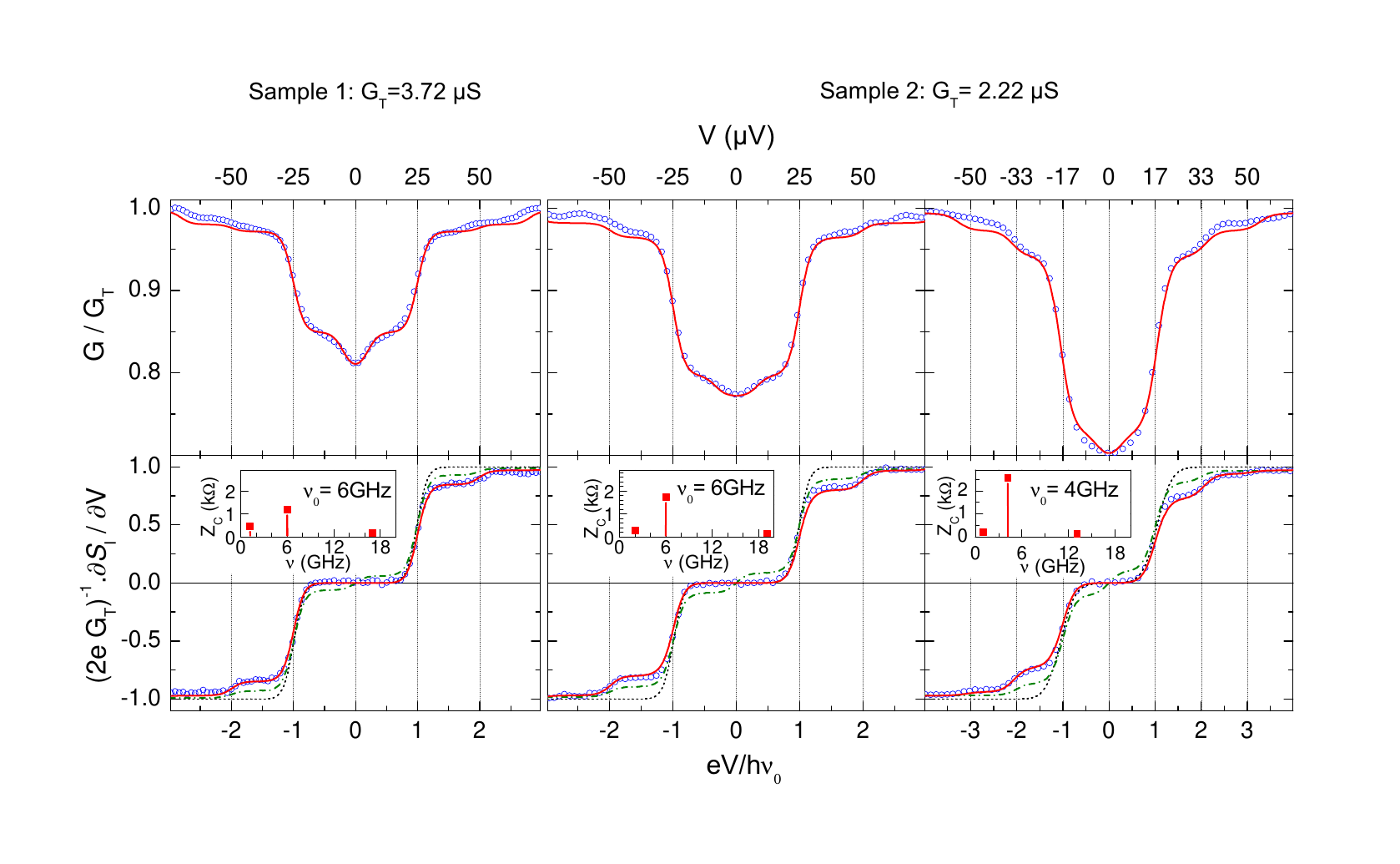}
 \caption{ Comparison between the measured conductance and noise blockade, and an extension of the dynamical Coulomb blockade theory. Normalized differential conductance $G ( V )$     (top) and current noise spectral density $\partial S_{\text{I}} ( \nu_{0} ,V) / \partial V$ (bottom). Open circles are experimental data measured at 15 mK. The left panel shows data measured on sample 1, with $\nu_{0}=6 \tmop{GHz}$, the center and right panel data measured on sample 2  with $\nu_{0} =6  \tmop{GHz}$ and 4 GHz, respectively. Solid red lines result from an analytical fit to the data involving series impedance made of three discrete modes shown in insets and the dotted black curve shows the non-interacting, finite frequency shot noise prediction. The green dot-dashed line represents the DCB expression for the current noise density symetrized with respect to frequency. }
\end{figure*}

Our experimental set-up is schematized in the lower panel of Fig. 1: a 100 $\times$ 100 nm$^2$ tunnel
junction with tunnel resistance $G^{-1}_{\text{T}}$ in the 100 k$\Omega$
range is embedded in an on-chip $\lambda /4$ coplanar resonator of resonant frequency$\nu_0$, whose inner
conductor is made of an array of identical and equally spaced Al/AlOx/Al
SQUIDs.  To a very good approximation, its lineic inductance is dominated by the Josephson inductance $L_{J} = \hbar (2eI_{0}
| \cos  (e \phi / \hbar )|a)^{-1}$, where $I_{0}$ is the maximum critical
current of one SQUID, $\phi$ the flux applied to each SQUID, and $a$ the
distance between adjacent SQUIDs. This increases the resonator impedance
$Z_{C}$ above 1 k$\Omega$, and allows to decrease $\nu_0$ while increasing $Z_C$  by applying a flux through the
SQUIDs. Two samples were fabricated and measured. In both cases the 6 GHz maximum frequency of the resonator ensures
$k_{\text{B}} T \ll h \nu_0$ at the 15 mK temperature of the experiment, so that thermal fluctuations do not blur Coulomb Blockade
effects.  The minimum zero flux lineic inductances of
the first and second resonators were designed at 8.80 $10^{-5}$
Hm\tmrsup{$-1$} and 3.95 $10^{-4}$ Hm\tmrsup{$-1$}, respectively.
 Note that Josephson transmission lines have been used to create non-linear resonators used as parametric 
amplifiers \cite{CastellanosNaturePhys.4.229}, or to probe how quantum phase slips 
drive them into an insulating state at $Z_C \gg R_K$ \cite{ChowPRL.81.207}.
We avoid this regime by keeping $Z_C$ in the k$\Omega$ range, which stills allows us to obtain sizeable DCB corrections. Keeping the current 
going through the resonator much smaller than $I_0$ ensures that the Josephson junctions can be considered as linear inductances.
The resistance $G^{-1}_{\text{T}}$
being much higher than $Z ( \nu )$, the impedance seen by one conduction
channel of the junction is not shunted by the parallel conductance of the
other channels {\cite{PhysRevLett.80.1956}}.
  The SQUIDs and the tunnel junction were fabricated  on a
Si/Si0$_{2}$ substrate using standard nanofabrication techniques \cite{SM}. In addition, a 30 $\times$ 50
$\times$ 0.3 $\mu$m\tmrsup{$3$} gold patch is inserted between the tunnel
junction and the SQUID array in order to evacuate the Joule power dissipated
at the tunnel junction via electron-phonon coupling. As an example, assuming a typical 
2nW $\mathrm{\mu m}^{-3} \mathrm{K}^{-5}$ electron-phonon coupling constant \cite{HuardPhysRevB}, a 100$\mu$V (resp. 1mV) bias on a 
200k$\mathrm{\Omega}$ tunnel resistance increases the electron temperature from 15mK to 
20mK (resp 50mK), keeping heating effects negligible. Note that the thermalization pad adds an additional 12fF to ground, which is taken into account to evaluate the total impedance seen by the tunnel junction. The chip is connected to
the biasing and measurement circuits through a commercial $50 \hspace{0.25em}
\Omega$ matched bias tee. The inductive (low frequency) path is used both to
bias the sample through a cold $13 \hspace{0.25em} \mathrm{M} \Omega$
resistor, and to measure the dc voltage across the tunnel junction and its
conductance $G(V)$. The capacitive (RF) path guides the radiation
$S_{\text{I}} (\nu ,V)$ emitted by the sample to a cryogenic isolator
anchored at 15 mK, to a cryogenic amplifier with a $\sim$2.5 K noise
temperature in the 4-8 GHz bandwidth, to room temperature band-pass filters,
and finally to a power "square law" detector, the output voltage of which is
proportional to its input microwave power. The isolator diverts the current noise of the amplifier to a $50 \Omega$ matched resistor
that re-emits to the sample a blackbody radiation only at the coldest
temperature, ensuring a negligible photon occupation of the resonator at GHz
frequencies. Finally, the signal $S_{I}(\nu ,V)$ is extracted from the large
noise floor of the cryogenic amplifier by a lock-in detection involving a 1
$\mu \text{V}$ sinusoidal modulation at 17 Hz on top of the dc voltage $V$.

We first characterized the on-chip microwave resonator by measuring the power emitted
by the electronic shot noise of the junction $S_{\Iota} \sim 2eI$ at high bias
voltage $V \sim 1 \tmop{mV}$ \cite{HofheinzPRL10}, where DCB effects are negligible. Under these
conditions, the spectral density of the emitted power is $2eV \mathrm{Re} [
\nobracket Z( \nu )] G_{\Tau} / |1+G_{\Tau} Z( \nu )|^{2} \simeq 2eVG_{\Tau}
\mathrm{Re} [Z( \nu )]$ since the tunnel resistance ($G^{-1}_{\text{T}}
=$230 k$\Omega$ /450 k$\Omega$ for sample 1/2) is much larger than the maximum detection impedance $Z ( \nu )$. This
spectral density is obtained using a heterodyne measurement implementing a 10
MHz-wide band pass filter at tunable frequency. As shown in Fig.2, the extracted $\mathrm{Re} [Z(
\nu )]$ is in satisfactory agreement with predictions. In particular, $Z ( \nu )$ shows the expected resonance,
with a resonant frequency $\nu_{0}$ decreasing with $\phi$, associated to an
increasing impedance and quality factor, which is limited by radiative losses. The maximum disagreement between the
measured maximum for $\mathrm{Re} [Z( \nu )]$ and the calculated one is about
15\%, which we attribute to an uncertainty in the calibration of the gain of the amplifying chain \cite{SM}. 
 We attribute the
additional structure around 5.7 GHz to a parasitic resonance in the detection
chain. Once our microwave environment calibrated, we measure both the
differential conductance $G ( V )$ of the tunnel junction and the voltage
derivative $\partial S_{I} ( \nu_{0} ,V) / \partial V$ of the noise emitted in
a 180 MHz bandwidth centered around the resonator frequency $\nu_{0}$, as a
function of the dc bias voltage applied to the junction. The conductance,
shown in the upper panel of Fig. 3 for both samples, is non-linear,
showing a stair-case behavior characteristic of DCB corrections due to a
single mode, rounded by the finite temperature {\cite{ingoldnazarov1992DCB}}.
The high characteristic impedance of our resonators yields DCB corrections to
the conductance 10 times higher than with standard microwave resonators
{\cite{HolstDCB1994,HofheinzPRL10}}, and shows not only the single photon
emission onset at bias voltage $V_{0} =h \nu_{0} /e$ but also the two photon
onset at $2V_{0}$. As shown in the lower panels of Fig. 3, the voltage derivative of emission noise power also 
displays a non-linear staircase shape, with
a first singularity at $V_{0}$, followed by a smaller step at $2V_{0}$. The
first step at $V_{0}$ is predicted in the standard -non including DCB effects-
finite frequency shot noise theory, represented by the dotted black curve in the
lower panel of Fig. 3. It has been observed in several experiments
{\cite{SchoelkopfDiffusivePRL1997,PhysRevLett.96.176601,ZakkaNoiseQPCPRL2007,PhysRevLett.99.206804}}
and can be dually understood either in terms of the finite time coherence of a
DC biased quantum conductor, or in terms of the energy cost of creating
excitations at frequency $\nu_0$  in the measuring apparatus
{\cite{LesovikLoosen,PhysRevB.62.R10637,PhysRevLett.84.1986}}. The second step
occurs at the onset voltage for the emission of two photons in the resonator
by a tunneling electron. The significant difference between the experimental
points and the non interacting prediction demonstrates the Coulomb Blockade of shot noise.

We now probe how the data shown in Fig. 3 can be quantitatively accounted for by
Eqs. 2 and 3, using our well controlled environment as an input to
evaluate $P(E)$. We model this environment as a series combination of three
discrete harmonic modes. The two higher frequency ones correspond to the
fundamental and first harmonic modes of the resonator. These two modes account with no
adjustable parameters for the observed variations above $V_{0} =h \nu_{0} /e$.
 Their characteristic impedance can be evaluated through
the standard formula 
$Z_C=\frac{2}{\nu_0 \mathrm{Im} Y'(\nu_0)}$, where $Y(\nu)$ is the environment's admittance, evaluated from our 
modeling of the Josephson transmission line. We  introduce an additional  lower frequency mode to account for the unexpected 3\% dip
in the differential conductance that we observe at low bias
voltage $|V| \lesssim 5 \mu V$. We attribute this low-frequency parasitic
resonance, which only slightly affects the data, to the bias-Tee.
The corresponding predictions, assuming an electron temperature $T_{e} =16$
mK corresponding to the temperature of the fridge's mixing chamber, are represented by the solid red curve in the
top graphs of Fig. 3. Note that at this
temperature, the $\sim 3.5k_{\text{B}} T/h \sim 1 \tmop{GHz}$ smearing
expected from the Fermi distribution is broader than the linewidth of the
modes of our resonator. This is why the discrete modes model, which yields
analytical expression for the $P (E)$ function {\cite{ingoldnazarov1992DCB}},
is able to reproduce the data. The emission noise data can be reproduced by Eq.
2 with excellent accuracy, whereas the expression corresponding to the
current noise spectral density symetrized with respect to frequency \cite{PhysRevB.53.7383},
$S^{\text{\tmop{sym}}}_{\Iota} ( \nu ,V) = [ S_{\Iota} (- \nu ,V)+S_{\Iota} (
\nu ,V) ] /2$, represented by the green dashed-dotted line in Fig.3, is not
compatible with our data\cite{SM}. Note that at low temperature $k_{\text{B}} T \ll h
\nu_{0}$ the relative size of the two-photon step is $\alpha /2$, which
explains why noise blockade is not seen with usual environment impedance
yielding values of $\alpha \sim 10^{-3}$. However, when considering the proposed primary shot noise thermometry \cite{Spietz20062003}, 
even such low values of $\alpha$ cause a systematic correction that should be considered to reach metrological
accuracy. Last, the data shown in the to left panels of Fig. 3 were taken for the maximum value of
$\nu_{0}$= 6 GHz. Applying a flux through the SQUIDs induces a stronger blockade due to
the increased Josephson inductance. As shown on the right panel of Fig. 3, the
higher impedance of sample 2 allows observing even three-photon processes when
the resonant frequency is set at 4 GHz (the lower end of our detection
bandwidth), yielding $Z_C \simeq$ 2.25 k$\Omega$ and a 30\%
reduction of the zero-bias conductance.

In conclusion, we have developed an original electromagnetic environment,
allowing to reach an unprecedented coupling between a quantum conductor and
a single mode environment. We took advantage of this to demonstrate the
Coulomb Blockade of the finite frequency noise of a tunnel junction. Two- and
three-photon processes are identified, in agreement with an extension to the
theory of Dynamical Coulomb Blockade. The experimental methods developed
here can be readily applied to quantum conductors of arbitrary transmissions,
for which a complete description of quantum transport in the presence of an
electromagnetic environment is still missing. Noticeably, they allow to probe
the Coulomb Blockade of shot noise in quantum point contacts
{\cite{Golubev01PRL86p4887,Yeyati01PRL87p046802,Cron01,AltimirasDCB2007,ParmentierStrongDCB2011,SouquetDCB2013}},
where DCB was recently demonstrated to bear a deep connection to the physics
of impurities in Luttinger liquids {\cite{JezouinStrongDCB2013}}, or quantum dots, where the
interplay between resonant tunneling through the dot and the coupling to the
environment was mapped to the physics of Majorana fermions
{\cite{MebrahtuNature488p61}}. This project was funded by the CNano-IDF
Shot-E-Phot and Masquel, the Triangle de la Physique DyCoBloS and ANR AnPhoTeQ grants.
Technical assistance from Patrice Jacques, Pierre-Fran{\c c}ois Orfila and
Pascal S{\'e}nat, as well as discussions within the Quantronics group, with
In{\`e}s Safi, Pascal Simon and Jean-Ren{\'e} Souquet are gratefully
acknowledged.

\end{document}


\title{Supplemental Material for the article ``Dynamical Coulomb Blockade of Shot
Noise''}

\author{Carles Altimiras}
\tmnote{Presently at NEST, Istituto Nanoscienze-CNR and Scuola Normale
Superiore, I-56127 Pisa, Italy}

\author{Olivier Parlavecchio}

\author{Philippe Joyez}

\author{Denis Vion}

\author{Patrice Roche}

\author{Daniel Esteve}

\author{Fabien Portier}
\affiliation{Service de Physique de l'Etat Condens{\'e} (CNRS URA 2464),
IRAMIS, CEA-Saclay, 91191 Gif-sur-Yvette, France}

\date{September 20, 2013\\
}

{\maketitle}

\section{Calculation of the current noise}

The circuit we consider consists of a pure tunnel element connected in series
with an arbitrary linear electromagnetic environment (See Fig. 1 of the main
text). In this picture, the capacitance of the real tunnel junction has been
incorporated into the impedance of the electromagnetic environment
{\cite{ingoldnazarov1992DCB}}. The Hamiltonian of the circuit is:
\[ H=H_{0} +H_{\text{T}} . \]
with
\[ H_{0} = \sum_{\ell} \varepsilon_{\ell} c_{\ell}^{\dagger} c_{\ell} +
   \sum_{r} \varepsilon_{r} c_{r}^{\dagger} c_{r} +H_{\text{\tmop{env}}} . \]
Here, the indexes $\ell$ and $r$ span all quasiparticle states in the left and
right electrodes, the $c_{\ell ,r}^{\dagger}$ (resp. $c_{\ell ,r}$) denote the
fermionic quasiparticle creation (resp. destruction) operators,
$H_{\text{\tmop{env}}}$ is the Hamiltonian of the electromagnetic environment,
and $H_{\text{T}} =T+T^{\dagger}$ is the tunneling Hamiltonian with $T$ and
$T^{\dagger}$ implementing the transfer of one electron across the barrier
from left to right and from right to left, respectiveley. This operator can be
decomposed as $T=e^{i \varphi} \Theta$ where the $e^{i \varphi}$ operator acts
only on the electromagnetic environment by translating the charge transfered
through the impedance by $e$, the charge of the tunneling electron, while
$\Theta = \sum_{\ell ,r} \tau_{\ell r} c_{\ell}^{\dagger} c_{r}$ acts only on
the quasiparticles in the electrodes. In this writing $\tau_{\ell r}$ is the
tunnel coupling amplitude of states $\ell$ and $r$, and $\varphi$ denotes the
phase difference operator across the tunnel element, acting on the
electromagnetic environment, and related to the voltage drop $V$ across the
tunnel element by $V= \frac{\hbar}{e} \frac{\partial \varphi}{\partial t}$.
The current operator through the tunnel element is given by $I=-
\frac{e}{\hbar} \frac{\partial H}{\partial \varphi} =-i \frac{e}{\hbar} (
T-T^{\dagger} )$.

The non-symmetrized noise current density $S_{\Iota} ( \nu )$ is the Fourier
transform of the current-current correlator:
\begin{equation}
  S_{\Iota} ( \nu ) =2 \int_{- \infty}^{\infty} \langle I(t)I(0) \rangle
  e^{-i2 \pi \nu t} dt \label{eq:SII}
\end{equation}
In this convention positive (resp. negative) frequencies correspond to energy
being emitted (resp. absorbed) by the quasiparticles to (resp. from) the
electromagnetic modes, and the current-current correlator function reads
\begin{eqnarray*}
  \langle I(t)I(0) \rangle & = & \frac{e^{2}}{\hbar^{2}} \{ \langle
  T(t)T^{\dagger} (0) \rangle + \langle T^{\dagger} (t)T(0) \rangle \} .
\end{eqnarray*}
We evaluate the averages by taking separate thermal equilibrium averages over
the unperturbed quasiparticles and environmental degrees of freedom:
\begin{equation}
\begin{split}
  \langle I(t)I(0) \rangle = \frac{e^{2}}{\hbar^{2}} \left\{ \langle e^{i \varphi
  (t)} e^{-i \varphi (0)} \rangle \langle \Theta (t) \Theta^{\dagger} (0)
  \rangle \right. \\ \left. + \langle e^{-i \varphi (t)} e^{i \varphi (0)} \rangle \langle
  \Theta^{\dagger} (t) \Theta (0) \rangle \right\} . \label{eq:curcor}
	\end{split}
\end{equation}
Unless at zero bias voltage where Eq. \ref{eq:curcor} is always exact, this
amounts to assuming that there exist unspecified relaxation mechanisms fast
enough to restore thermal equilibrium both in the electrodes and in the
impedance between tunneling events. The validity of this latter assumption is
highly dependent on the details of the system and should be checked on a case
by case basis.

Within this assumption, the occupation probability of the quasiparticle energy
levels in the electrodes is given by the Fermi function $f ( \epsilon )$ at
the inverse temperature $\beta$. We further assume constant densities of
states $\rho_{l} , \rho_{r}$ in the electrodes and we replace the $|
\tau_{\ell r} |^{2}$ by their average value $| \tau |^{2}$ over all $l,r$
states. The average over quasiparticles degrees of freedom thus takes the form
\begin{eqnarray*}
  \langle \Theta (t) \Theta^{\dagger} (0) \rangle & = & \left\langle
  \sum_{\ell ,r} | \tau_{\ell r} |^{2} c_{\ell}^{\dagger} (t)c_{\ell} (0)c_{r}
  (t)c_{r}^{\dagger} (0) \right\rangle\\
  & = & | \tau |^{2} \rho_{l} \rho_{r}   \int d \varepsilon d \varepsilon' f
  ( \varepsilon ) ( 1-f( \varepsilon + \varepsilon' ) ) e^{-i \varepsilon' t/
  \hbar}\\
  & = & \frac{\hbar G_{T}}{2 \pi e^{2}} \int d \varepsilon' \gamma (
  \varepsilon' ) e^{-i \varepsilon' t/ \hbar}\\
  & = & \frac{\hbar G_{T}}{2 \pi e^{2}} \tilde{\gamma} (t) ,
\end{eqnarray*}
where $G_{T} =4 \pi^{2} G_{K} | \tau |^{2} \rho_{l} \rho_{r}$ denotes the
tunnel conductance of the junction, which we assume much smaller than $G_{K}
=e^{2} /h$, allowing to treat $H_{T}$ at the lowest order in perturbation
theory; $\tilde{\gamma} (t)$ denotes the inverse Fourier transform of
\[ \begin{array}{rcl}
     \gamma ( \epsilon ) & = & \int d \varepsilon' f ( \varepsilon' ) ( 1-f(
     \varepsilon' + \varepsilon ) )\\
     & = & \frac{\epsilon}{1-e^{- \beta \epsilon}}\\
     & = & \frac{\epsilon}{2} \left( 1+ \coth \frac{\beta \epsilon}{2}
     \right)\\
     & = & n_{B} ( | \epsilon | ) + \epsilon \hspace{0.25em} \theta (
     \epsilon ) ,
   \end{array} \]
where $n_{B}$ is the Bose distribution function. The function $\gamma (
\epsilon )$ is proportional to the number of possible electron-hole
excitations in the Fermi seas of the electrodes with an energy difference of
$\epsilon$ between the final and initial states. The same function is
encountered in other contexts, such as in the noise of a resistor, for
emission and absorption by bosonic degrees of freedom (see below). Even though
$\gamma ( \epsilon )$ is not bounded, its inverse Fourier transform can
nevertheless be expressed using distribution functions, meant to be integrated
with proper functions {\cite{PhysRevB.56.1848,odintsov_effect_1989-1}}:
\begin{equation}
  \tilde{\gamma} (t) =i \pi \hbar^{2} \frac{d}{dt} \delta (t) -
  \frac{\pi^{2}}{\beta^{2}} \sinh^{-2} \frac{\pi t}{\hbar \beta} .
  \label{eq:gammatilde}
\end{equation}
The second quasiparticle term $\langle \Theta^{\dagger} (t) \Theta (0)
\rangle$ can be checked to give the same result, i.e. $\langle
\Theta^{\dagger} (t) \Theta (0) \rangle = \langle \Theta (t) \Theta^{\dagger}
(0) \rangle$.

To evaluate the average over the electromagnetic degrees of freedom, we
decompose the phase difference across the tunnel element $\varphi (t) =
\varphi_{0} (t) + \tilde{\varphi} (t)$ into the deterministic part
$\varphi_{0} (t) =eVt/ \hbar$ corresponding to the dc voltage $V$ across the
junction, and a fluctuating random phase $\tilde{\varphi} (t)$ caused by the
fluctuations in the electromagnetic environment. For a linear electromagnetic
environment at equilibrium, phase fluctuations are Gaussian. In this case, the
averages we need to evaluate in Supplemental Material Eq. \ref{eq:curcor} \
can be expressed as $\langle \exp \pm i \tilde{\varphi} (t) \exp \mp i
\tilde{\varphi} (0) \rangle = \exp  J (t)$, where $J (t) = \langle (
\tilde{\varphi} (t)- \tilde{\varphi} (0) ) \tilde{\varphi} (0) \rangle$ is the
phase-phase correlation function {\cite{ingoldnazarov1992DCB}}. The phase
being proportional to the time derivative of the voltage, the phase-phase
correlation function is related to the voltage noise on impedance seen by the
junction, which can be obtained by the quantum fluctuation-dissipation theorem
{\cite{ingoldnazarov1992DCB}}:
\begin{equation}
  J (t) =h R^{^{-1}}_{K} \int_{- \infty}^{+ \infty} \frac{S_{\text{V \text{}}}
  ( \nu )}{(h \nu )^{2}} (e^{-i2 \pi \nu t} -1) d \nu , \label{eq:j(t)}
\end{equation}
with
\[ S_{\text{V}} ( \nu ) =2 \text{\tmop{Re}} Z ( \nu ) \gamma (h \nu ) \]
being the (equilibrium) non-symmetrized voltage noise across the environment
impedance {\cite{RevModPhys.82.1155}}. Collecting the above terms in
Supplemental Material Eq. \ref{eq:curcor}, we obtain the final expression for
the current-current correlator:
\[ \langle I(t)I(0) \rangle = \frac{G_{\text{T}}}{2 \pi \hbar}  2  \cos
   \frac{eVt}{\hbar}  e^{J (t)} \tilde{\gamma} (t) . \]
Inserting this result in the Fourier transform (\ref{eq:SII}), the current
noise can be expressed as
\[ S_{\Iota} ( \nu ) =2G_{\text{T}} [ \gamma \ast P(-h \nu +eV)+ \gamma \ast
   P(-h \nu -eV) ] , \]
where * denotes convolution, and $P (E) = \frac{1}{2 \pi \hbar} \int_{-
\infty}^{\infty} e^{J (t) +iEt/ \hbar} \text{\tmem{\tmop{dt}} \tmem{}}$ is the
Fourier transform of $\exp [J(t)]$. This function $P (E)$ is interpreted as
the probability for the environment to absorb the algebraical energy $E$
during a tunnel event, and the convolution product:
\[ \gamma \ast P (E) = \int d \varepsilon' \gamma ( \varepsilon' ) P (E-
   \epsilon' ) \]
simply accounts for all the possible ways to split an energy $E$ between the
quasiparticles degrees of freedom and the environment modes. For the case of
an environment of vanishing impedance, $P (E) \rightarrow \delta (E)$, and we
recover a well known result
\[ S_{\text{I}} ( \nu ) \rightarrow S_{\text{I} } ( \nu ) =2 G_{\text{T}} [
   \gamma (-h \nu +eV)+ \gamma (-h \nu -eV) ] . \]

\section{Sample Fabrication}

The 300 nm thick gold ground
plane of the resonator and thermalization pad were obtained by optical lithography, followed by evaporation 
and lift-off. 
SQUIDs
where fabricated following the process described in Ref.
\cite{PopReproducibleJunctionsArXiv2012}: the SQUIDs
(see the top inset) are obtained by double angle deposition of
($20/40\,\mathrm{nm}$) thin aluminum electrodes, with a $20'$
oxydation of the first electrode at $400\,\mathrm{mBar}$ of a ($85\%\,\mathrm{O}_2/15\%\,\mathrm{Ar}$) mixture. 
Before the evaporation, the substrate was cleaned by rinsing in ethanol and Reactive Ion Etching in an oxygen plasma \cite{Plasma}. The normal junction was
obtained using the same technique, with 30/60 nm thick copper electrodes and
an aluminum oxyde tunnel barrier (5 nm thick aluminum oxidized for 15 minutes
at a 800 mBar (85\%O$_{2}$, 15\%Ar) mixture). 

\section{Details on the Josephson transmission line}

For sample 1 (resp. 2), our resonator consists in a $360\,\mathrm{\mu m}$ (resp.  $91\,\mathrm{\mu m}$)  long
Josephson metamaterial line containing 72 (resp. 38) lithographically identical and
evenly spaced SQUIDs with a $5\,\mathrm{\mu m}$  (resp.  $2.4\,\mathrm{\mu m}$) period.   
The SQUIDs tunnel barriers of sample 1 (resp. sample 2) have an area of
$0.5\,\mathrm{\mu m^2}$ (resp. $0.5\,\mathrm{\mu m^2}$) each resulting in a room temperature
tunnel resistance $R_\mathrm{N}=720\,\Omega$ (resp. $R_\mathrm{N}=1880 \,\Omega$). To assess that the SQUIDs in the array are identical, we have performed reproducibility tests,
yielding constant values of $R_\mathrm{N}$ (within  a few $\%$) over millimetric distances. Assuming a superconducting gap $\Delta=180\,\mathrm{\mu
  eV}$ and a 17\% increase of the tunnel resistance between room temperature and base temperature \cite{GloosGTTunnelAPL}, one obtains a zero flux critical current for the SQUIDs $I_C= 671$nA for sample 1 and $I_C= $268 nA, 
  corresponding to $L_J(\phi=0)=0.49$ nH for sample 1 and $L_J(\phi=0)=1.25$ nH for sample 2. 
 This corresponds to an effective lineic inductance   $\mathcal{L}\simeq\mathrm{100\,\mu H.m^{-1}}$ at zero
magnetic flux and frequency much lower than the Josephson plasma frequency of the junctions $\nu_\textrm{P}$\cite{JosephsonRMP1964}. Assuming  a capacitance for the junctions of the order of 80 fF/$\mu$m$^2$ yields $\nu_\textrm{P}\simeq$ 25 GHz.
Note that our simple fabrication mask produces 10 times bigger
Josephson junction in between adjacent SQUIDs, resulting in an additional $\sim \mathcal{L}\simeq\mathrm{10\,\mu H.m^{-1}}$ lineic  inductance for sample 1 and $\sim \mathcal{L}\simeq\mathrm{25\,\mu H.m^{-1}}$ for sample 2.
The  $\sim \mathcal {L}\simeq\mathrm{1\,\mu H.m^{-1}}$ electromagnetic inductance associated to our geometry is negligible. 
  With the designed lineic capacitance
$\mathcal{C}=75 \,\mathrm{pF.m^{-1}}$, the length of the resonator
sets the first resonance at $\nu_0\simeq 8\,\mathrm{GHz}$ for sample 1 and $\nu_0\simeq 12\,\mathrm{GHz}$ for sample 2. 
The 12 fF shunting capacitance of the themalization pad reduces these frequencies to $\nu_0 \simeq 6\,\mathrm{GHz}$ in both cases.

\section{Details on setup and calibration}

We describe the calibration of the low frequency circuitry for voltage bias
and current measurement as well the microwave components used to define the
environment of the junction and to measure the emitted radiation.

\subsection{Low frequency circuit}

In addition to the components depicted in Fig.~1 of the main text, the low
frequency circuit includes a copper powder filter anchored on the mixing
chamber, as well as a distributed RC filter made with a resistive wire (50 cm
of IsaOhm 304 $\Omega$ m$^{-1}$) winded around a copper rod, and glued with
silver epoxy on a copper plate in good thermal contact with the mixing
chamber. Both are inserted between the 13 M$\Omega$ bias resistor and the bias
T and are represented by the 170$  \Omega$/450 pF RC filter on the biasing
line in Supplemental Material Fig. 1. The distributed RC filter has two
benefits on the effective electron temperature of our experiment: it provides
a high frequency filtering that reduces the polarisation noise as well as
thermalisation of the electrons. The copper powder filter is meant to absorb
parasitic microwave noise. The line allowing to measure the low frequency
response of the junction is filtered by a multipole RC low pass filter, made
with a succession of 2 k$\Omega$ Nickel-Chromium resistances and 1 nF
capacitances to ground. The NiCr resistances were checked in an independent
cool-down to change by less than 1\%, which allows to calibrate the 13
M$\Omega$ resitor in-situ, with a precision better than 1\%, which in turn
allows us to determine the dc voltage $V$ applied to the tunnel junction. The
validity of this callibrattion is confirmed by the quality of the comparison
between the observed steps in $\partial  S_{I} ( \nu ) / \partial  V$ and our
predictions.

\subsection{Microwave circuit and calibration}

The microwave chain comprises a bias T, two 4-8 GHz cryogenic circulators
anchored at the mixing chamber, as well as a 4-8 GHz bandpass filter and a 12
GHz low pass gaussian absorptive filter (see Supplemental Material Fig. 2).
These elements are anchored on the mixing chamber and are meant to protect the
sample from the back-action noise of the amplifier. The quantitative
determination of the detection impedance relies on the detection of the power
emitted by the shot noise of the tunnel junction in the high bias regime. We
bias the junction at $\sim 1  \tmop{mV}$, where DCB corrections are
negligible, \ so that $S_{I} =2eI$ at frequencies $| \nu | \ll eV/h \simeq 0.5
\tmop{THz}$. In order to separate this noise from the noise floor of the
cryogenic amplifier, we then apply small variations of the bias voltage and
measure the corresponding changes in the measured microwave power with a
lock-in amplifier. The conversion of $S_{I}$ into emitted microwave power
depends on the environment impedance $Z ( \nu )$ seen by the tunneling
resistance $R_{T}$. First, only a fraction $R_{T}^{2} / |R_{T} +Z( \nu )|^{2}$
of the current noise is absorbed by the environment. The current noise in the
environment has then to be multiplied by $\tmop{Re}   [ Z( \nu ) ]$ to obtain
the microwave power emitted by the electronic shot noise:
\begin{equation}
  S_{P} ( \nu ) =2eV \frac{\tmop{Re}   [ Z( \nu ) ] G_{\Tau}^{-1}}{|Z( \nu
  )+G_{\Tau}^{-1} |^{2}} \simeq 2eV \frac{\tmop{Re}   [ Z( \nu ) ]
  G_{\Tau}^{-1}}{[ \tmop{Re}   [ Z( \nu ) ] +G_{\Tau}^{-1} ]^{2}} .
  \label{eq:shotnoise}
\end{equation}
The last approximation, $\tmop{Im}   [ Z( \nu ) ] \ll G_{\Tau}^{-1}$ is
satisfied with a precision better than 2\%. Finally, what is actually detected
at room temperature is the amplified microwave power:
\begin{equation}
  S^{\tmop{RT}}_{P} ( \nu ) =2eV G ( \nu ) \frac{\tmop{Re}   [ Z( \nu ) ]
  G_{\Tau}^{-1}}{[ \tmop{Re}   [ Z( \nu ) ] +G_{\Tau}^{-1} ]^{2}} .
  \label{eq:shotnoiseamplified}
\end{equation}
Supplemental Material Eq. \ref{eq:shotnoiseamplified} shows that the extracted
$\tmop{Re} [ Z( \nu ) ]$ depends on the gain of the microwave chain $G ( \nu
)$, which has to be determined in-situ and independently. To do so, we
inserted a 20 dB directional coupler between the sample and the bias Tee, and
injected through an independently calibrated injection line, comprising 70 dB
attenuation distributed between 4.2 K and the mixing chamber temperature (see
Supplemental Material Fig. 2). Both the attenuators and the directional
coupler were calibrated at 4.2 K. The resonator is tuned at a resonance well
below \ 4 GHz, so that the entire microwave power is reflected by the sample,
allowing to calibrate the gain of the detection chain, albeit doubling the
insertion losses of the bias Tee and of the 10 cm microwave cable connecting
it to the sample. However, this parasitic contribution can be substracted
thanks to independent calibrations. Due to lack of space, we then had to
remove one of the circulators for this experiment, which increased the
electronic temperature from 16 mK to 25 mK, which explains that we calibrated
the detection impedance in an independant run of the experiment. 

\section{Extracting the current noise}

We discuss here the possible consequences of the fact that the detection
impedance is not negligible compared to the tunneling resistance. More
specifically, we show that due to the variations of the tunneling resistance
with bias voltage, measuring $\partial  S_{P} ( \nu ) / \partial  V$ is not
rigourously equivalent to measuring $\partial  S_{I} / \partial  V$. However,
the error introduced by this approximation can be shown to be negligible.

Due to the non linearity of the tunnel transfer, the power emitted by the
junction biased at bias $V  \tmop{reads}$
\begin{equation}
  S_{P} ( \nu ) = \tmop{Re}   [ Z( \nu ) ] \left| \frac{Z_{\text{T}} ( \nu ,V
  )}{Z( \nu )+Z_{\text{T}} ( \nu ,V )^{}} \right|^{2} S_{I} ( V, \nu )
  \label{eq:sp} .
\end{equation}
Here $Z_{\text{T}} ( \nu ,V )$ is the differential impedance of the junction,
biased at voltage $V$, at the measurement frequency $\nu$. Supplemental
Material Eq. \ref{eq:sp} is valid as long as the ac current going through the
junction as a consequence of the shot noise is small enough for the response
of the junction $Z_{\text{T}} ( \nu ,V )$ to remain in the linear regime. In
that case, the modulation of the output voltage of the quadractic detector
that we measure is proportional to
\begin{equation}
\begin{split}
  \frac{\partial  S_{P} ( \nu )}{\partial  V} = \tmop{Re}   [ Z( \nu ) ]
  \left[ \left| \frac{Z_{\text{T}} ( \nu ,V )}{Z( \nu )+Z_{\text{T}} ( \nu ,V
  )^{}} \right|^{2} \frac{\partial S_{I} ( V, \nu )}{\partial  V} \right.\\ \left. +S_{I} ( V,
  \nu ) \frac{\partial}{\partial V} \left| \frac{Z_{\text{T}} ( \nu ,V )}{Z(
  \nu )+Z_{\text{T}} ( \nu ,V )^{}} \right|^{2} \right] \label{eq:dspdv} .
	\end{split}
\end{equation}

\ We can deduce the expected variations of $Z_{\text{T}} ( \nu ,V )$ with
bias voltage from the dc transport properties of the junction via
\begin{equation}
  Z^{-1}_{\text{T}} ( \nu ,V ) = \frac{e}{2 h  \nu} [ I ( V+h \nu /e ) -I (
  V-h \nu /e ) ] \label{eq:invzt} .
\end{equation}
Inserting Eq. \ref{eq:invzt} in Eq. \ref{eq:dspdv} shows that the associated
corrections are negligeable, so that detecting $\frac{\partial  S_{P} ( \nu
)}{\partial  V}$ gives direct access to $\frac{\partial S_{I} ( V, \nu
)}{\partial  V}$ within a precision better than 1\%.

\begin{figure*}[h]

  \resizebox{10cm}{!}{\includegraphics{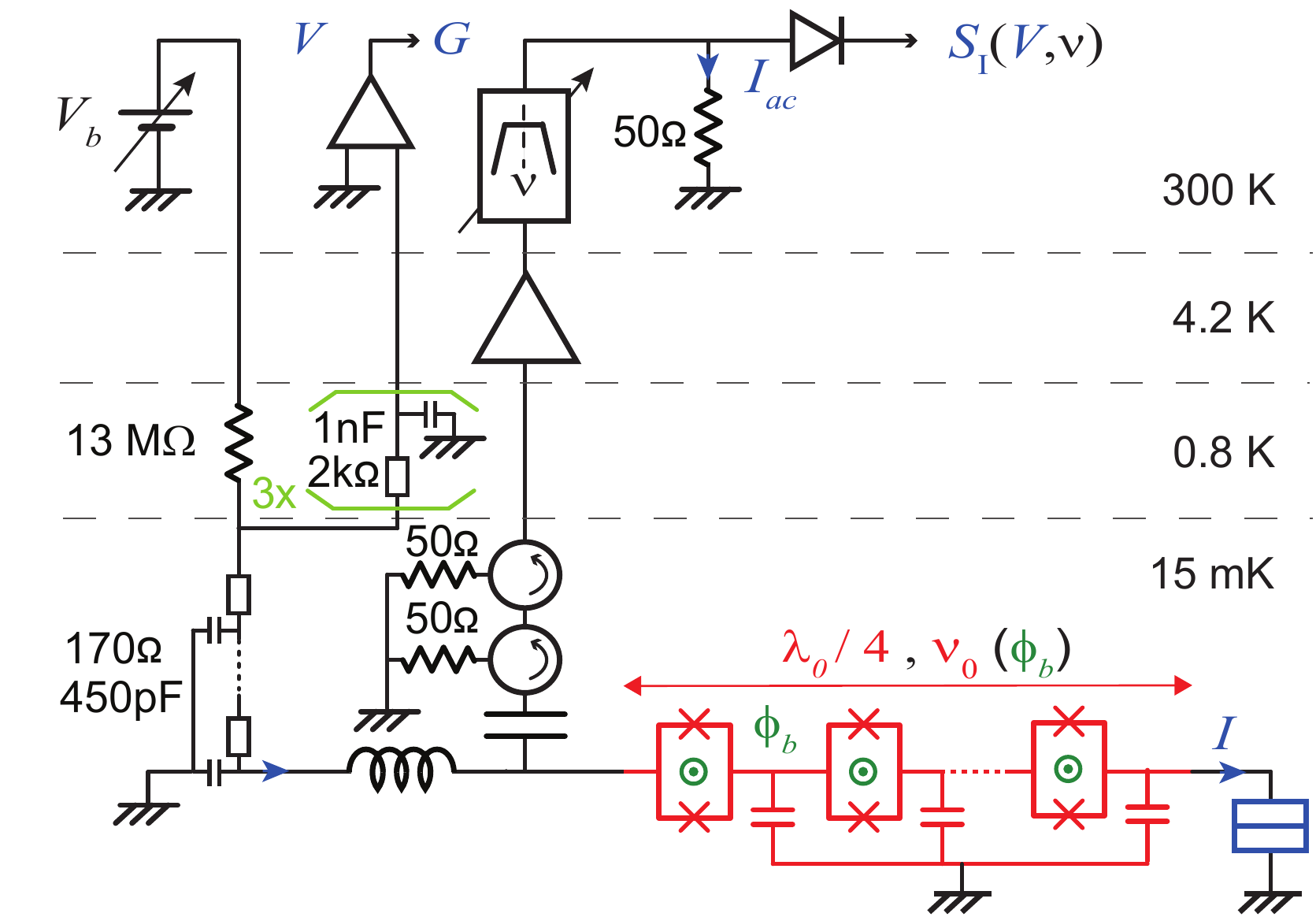}}
  \caption{More detailled view of the experimental setup. Only circuit
  components inside the refrigerator are shown in full detail.}
\end{figure*}

\begin{figure*}[h]
  \resizebox{10cm}{!}{\includegraphics{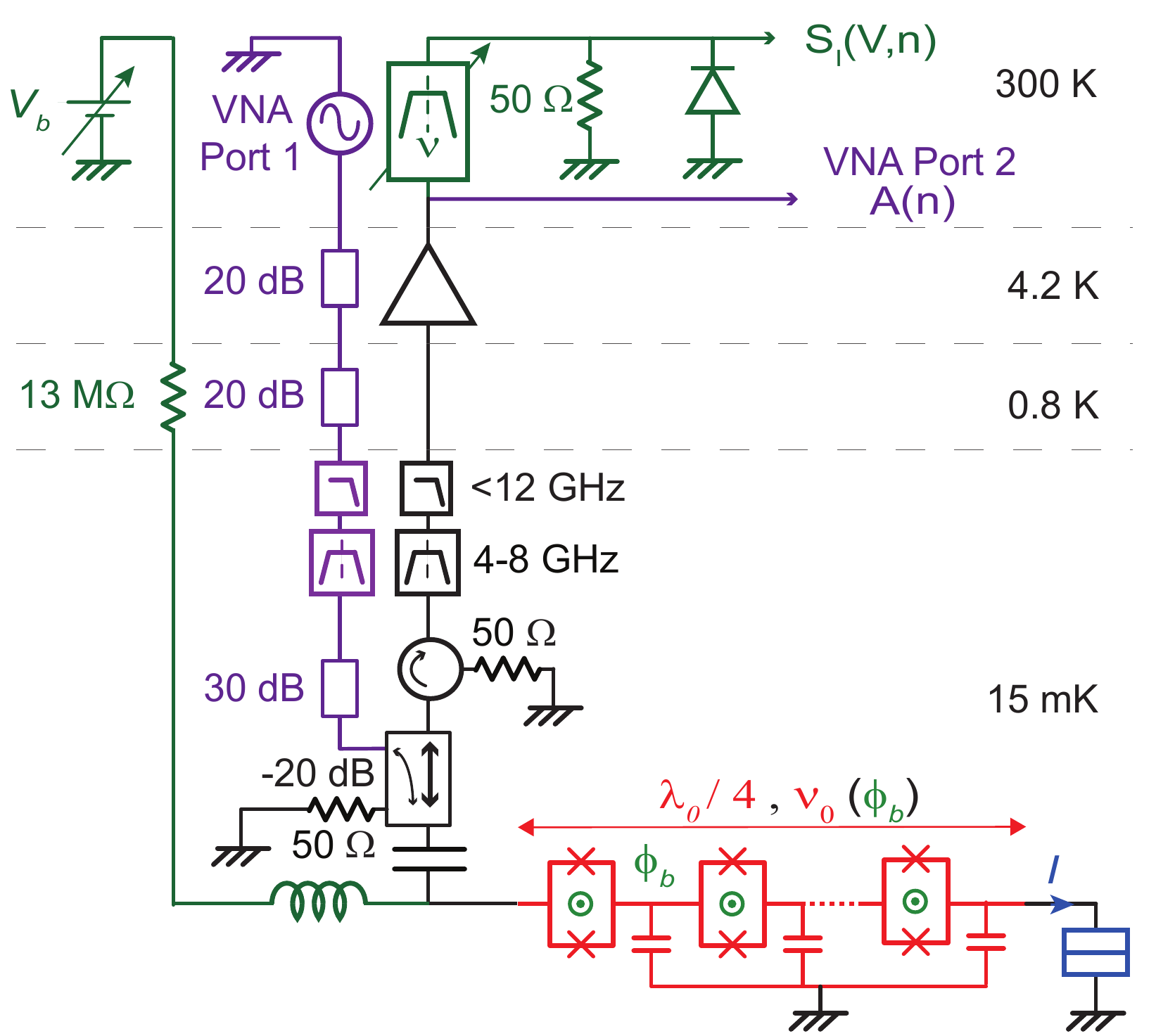}}
  \caption{Characterization of the environment impedance: the microwave chain
  is calibrated using a vectorial network analyzer, setting the resonance
  frequency well below 4 GHz. }
\end{figure*}

\section{Note on noise symetrization}

Let us  add a note about the symetrization of the correlator being measured in our high frequency noise measurements.
On one hand the microwave amplifier probes its input voltage \cite{RevModPhys.82.1155}, which can be written as 

\begin{equation}
V_{\mathrm{in}}(t)=-\sqrt{\frac{\hbar Z_0}{4\pi}}\int_B  \omega^{1/2}(i a_{\mathrm{in}} (\omega) e^{-i \omega (t - x/c)} + \mathrm{h.c.}) d\omega,
\end{equation}

where $a_{\mathrm{in}}(\omega)$ stands for the destruction operator of a input (right moving) photon at frequency $\omega$, $c$ is the velocity of electromagnetic waves in the 
transmission line connecting the sample to the amplifier, of characteristic impedance $Z_0 = 50 \Omega$, and $B$ the measurement bandwidth. The output voltage reads:

\begin{equation}
V_{\mathrm{out}}(t)=-\sqrt{\frac{\hbar Z_0}{4\pi}}\int_B  \omega^{1/2}(i a_{\mathrm{out}} (\omega) e^{-i \omega (t - x/c)} + \mathrm{h.c.}) d\omega,
\end{equation}

with

\begin{equation}
a_{\mathrm{out}}=\sqrt{G} a_{\mathrm{in}} + \sqrt{G-1} f^\dagger
\end{equation}

with $G$ the gain of the amplifier and $f^\dagger$ representing the amplifier's noise  \cite{RevModPhys.82.1155}. Thus the final power measurement $\propto <V_{\mathrm{out}}^2>$
contains, on top of the amplifier's noise, a term proportional to

\begin{eqnarray}
\left\langle  V^2_{\mathrm{in}}\right\rangle&=&\frac{\hbar Z_0}{4\pi}\int_B  \omega  \left\langle a_{\mathrm{in}}(\omega) a^\dagger_{\mathrm{in}}(\omega) + a^\dagger_{\mathrm{in}}(\omega) a_{\mathrm{in}}(\omega) \right\rangle d\omega \\
&=&\frac{\hbar Z_0}{2\pi}\int_B  \omega  \left[ \left\langle a^\dagger_{\mathrm{in}}(\omega) a_{\mathrm{in}}(\omega))\right\rangle + \frac{1}{2} \right]   d\omega, 
\end{eqnarray}

thus containing a power representing the zero point motion of the line, associated to a power $\hbar \omega /2$ per unit bandwidth. In other words, the amplifier gives access to the sum of the absorption 
and emission noise of the input line, unlike quantum detectors \cite{PhysRevLett.105.166801,PhysRevB.85.085435} which allow detecting them separately.
However, this does not imply that we measure the \emph{electronic current noise of the sample} symetrized with respect to frequency:
In our experiment, we measure the excess output noise power associated to the dc biasing of our sample. This results in a excess population of the 
input field  $<a^\dagger_{\mathrm{in}}(\omega) a_{\mathrm{in}}(\omega)>$, which is itself proportional to the photons leakage rate out of our 
resonator. Thanks to the circulators anchored at the mixing chamber temperature $T \ll h \nu_0 /k_\mathrm{B}$, the latter is kept 
very close to its ground state (the maximum average number of photons in the resonator is below 5\% for the experiments reported in the paper), and can thus only absorb energy from the electronic 
current fluctuations (in other words $P(h\nu) \ll P(-h\nu)$). This is why our signal is proportional to the emission noise power  $ S_{I}(\nu ,V)$ and not the noise power symetrized with respect to frequency $S^{\text{\tmop{sym}}}_{\Iota}(\nu ,V) = [S_{I} (- \nu ,V)+S_{I} (\nu ,V) ] /2$, which also contains the electronic zero point motion.